\documentclass{article}

\usepackage{graphicx}  
\usepackage{amsmath}   
\usepackage{amssymb}   

\def\eq#1{{Eq.~(\ref{#1})}}

\newcommand{\cc}{cosmological constant}

\newcommand{\Cal}[1]{\ensuremath{\mathcal{#1}}}

\newcommand{\LL}{Lanczos-Lovelock }
\newcommand{\D}{\ensuremath{\nabla}}
\newcommand{\Riem}[4]{\ensuremath{R^{#1 #2}_{#3 #4}}}
\newcommand{\Alt}[6]{\ensuremath{\delta^{#1 #2 ... #3}_{#4 #5
      ... #6}}} 

\newcommand{\AltC}[8]{\ensuremath{\delta^{#1 #2 #3... #4}_{#5 #6 #7
      ... #8}}}  
\newcommand{\sD}[1]{\sum_{m=1}^{K}{#1}}
\newcommand{\LDm}{\ensuremath{\Cal{L}^{(D)}_m}}
\newcommand{\eqn}[1]{Eq.\eqref{#1}}
\newcommand{\ph}[1]{\phantom{#1}}

\def\frab#1#2{\left(\frac{#1}{#2}\right)} 

\title{Emergent gravity and Dark Energy}
\author{T. Padmanabhan\\
IUCAA, Pune University Campus, \\
Ganeshkhind, Pune 411 007, INDIA\\
email: nabhan@iucaa.ernet.in}
\date{ }

\begin{document}

\maketitle

\section{The rise of the Dark Energy}

Given the expansion rate of the universe in terms of the Hubble constant  $H_0=(\dot a/a)_0$, one can define  a \textit{critical energy density} $\rho_c=3H^2_0/8\pi G$ which is required to make the spatial sections of the universe compact. It is convenient to measure the 
energy densities of the different species, which drive the expansion of the universe, in terms of this critical density using the dimensionless parameters $\Omega_i=\rho_i/\rho_c$ (with $i$ denoting
the different components like baryons, dark matter, radiation, etc.) The simplest possible universe one could imagine would have just baryons and radiation. However, host of astronomical observations available since mid-70s indicated that the bulk of the matter in the universe is nonbaryonic and dark. Around the same time, the theoretical prejudice for $\Omega_{tot}=1$ gained momentum, largely led by the inflationary paradigm. During the eighties, this led many theoreticians to push (wrongly!) for a model of the with $\Omega_{tot}\approx\Omega_{DM}\approx1$ in spite of the fact that host of astronomical observations demanded that $\Omega_{DM}\simeq 0.2-0.3$.

The indications that the universe indeed has another component of energy density started accumulating in the late eighties and early nineties. Early analysis of several observations\cite{earlyde} indicated that this component is unclustered and has negative pressure. This is confirmed dramatically by the supernova observations in the late nineties (see Ref.~\cite{sn}; for a critical look at the current data, see Ref.~\cite{tptirthsn1}).  The observations suggest that the missing component has 
$w=p/\rho\lesssim-0.78$
and contributes $\Omega_{DE}\cong 0.60-0.75$. 

The simplest choice for such \textit{dark energy} with negative pressure is the cosmological constant which is  a term that can be added to Einstein's equations. This term acts like  a fluid with an equation of state $p_{DE}=-\rho_{DE}$.
Combining this with all other observations \cite{cmbr,baryon,h}, we end up with a weird  composition for the universe with
 $0.98\lesssim\Omega_{tot}\lesssim 1.08$  in which radiation (R), baryons (B), dark matter, made of weakly interacting massive particles (DM) and dark energy (DE) contributes  $\Omega_R\simeq 5\times 10^{-5},\Omega_B\simeq 0.04,\Omega_{DM}\simeq 0.26,\Omega_{DE}\simeq 0.7,$ respectively. So the bulk of the energy density in the universe is contributed by \textit{dark energy}, which is theme of this article.

The  remarkably successful paradigm of conventional cosmology is based on these numbers and works \cite{adcos} as follows:
The  key idea is that if there existed small fluctuations in the energy density in the early universe, then gravitational instability can amplify them   leading to structures like galaxies etc. today. The popular procedure for generating these fluctuations is based on the idea that if the very early universe went through an inflationary phase \cite{inflation}, then the quantum fluctuations of the field driving the inflation can lead to energy density fluctuations \cite{genofpert,tplp}. While the inflationary models are far from unique and hence lacks predictive power, it is certainly possible to construct models of inflation such that these fluctuations are described by a Gaussian random field and are characterized by a power spectrum of the form $P(k)=A k^n$ with $n\simeq 1$. The inflationary models cannot predict the value of the amplitude $A$ in an unambiguous manner. But it can be determined from CMBR observations and the inflationary model parameters can be fine-tuned to reproduce the observed value. The CMBR observations are consistent with the inflationary model for the generation of perturbations and gives $A\simeq (28.3 h^{-1} Mpc)^4$ and $n\lesssim 1$. (The first results were from COBE \cite{cobeanaly} and
WMAP has re-confirmed them with far greater accuracy). One can evolve the initial perturbations by linear perturbation theory
when the perturbation is small. But when $\delta\approx(\delta\rho/\rho)$ is comparable to unity the perturbation theory
breaks down and one has to resort to numerical  simulations \cite{baryonsimulations}
 or theoretical models based on approximate 
 ansatz \cite{nlapprox, nsr} to understand their evolution --- especially the baryonic part, that leads to observed structures in the universe. This rapid summary shows that modeling the universe and comparing the theory with observations is a rather involved affair; but the results obtained from all these
 attempts are broadly consistent with observations.

 To the zeroth order, the universe is characterized by just seven numbers: $h\approx 0.7$ describing the current rate of expansion; $\Omega_{DE}\simeq 0.7,\Omega_{DM}\simeq 0.26,\Omega_B\simeq 0.04,\Omega_R\simeq 5\times 10^{-5}$ giving the composition of the universe; the amplitude $A\simeq (28.3 h^{-1} Mpc)^4$ and the index $n\simeq 1$ of the initial perturbations. 
 
 \section{A first look at \cc\ and its problems}
 
 The remaining challenge, of course,  is to make some sense out of these numbers themselves from a more fundamental point of view. Among all these components, the dark energy, which exerts negative pressure, is probably the weirdest one and has attracted most of the attention.

 The key observational feature of dark energy is that, when treated as a fluid with a stress tensor $T^a_b=$ dia     $(\rho, -p, -p,-p)$, 
it has an equation state $p=w\rho$ with $w \lesssim -0.8$ at the present epoch. 
The spatial part  ${\bf g}$  of the geodesic acceleration (which measures the 
  relative acceleration of two geodesics in the spacetime) satisfies an \textit{exact} equation
  in general relativity  given by:
  \begin{equation}
  \nabla \cdot {\bf g} = - 4\pi G (\rho + 3p)
  \label{nextnine}
  \end{equation} 
 This  shows that the source of geodesic  acceleration is $(\rho + 3p)$ and not $\rho$.
  As long as $(\rho + 3p) > 0$, gravity remains attractive while $(\rho + 3p) <0$ can
  lead to `repulsive' gravitational effects. In other words, dark energy with sufficiently negative pressure will
  accelerate the expansion of the universe, once it starts dominating over the normal matter.  This is precisely what is established from the study of high redshift supernova, which can be used to determine the expansion
rate of the universe in the past \cite{sn,snls}.

The simplest model for  a fluid with negative pressure is not a fluid at all but the
cosmological constant with $w=-1,\rho =-p=$ constant (for a few of the recent reviews, see ref. \cite{cc}).  The cosmological constant  introduces a fundamental length scale in the theory $L_\Lambda\equiv H_\Lambda^{-1}$, related to the constant dark energy density $\rho_{_{\rm DE}}$ by 
$H_\Lambda^2\equiv (8\pi G\rho_{_{\rm DE}}/3)$.
Though, in classical general relativity,
    based on  $G, c $ and $L_\Lambda$,  it
  is not possible to construct any dimensionless combination from these constants, when one introduces the Planck constant, $\hbar$, it is  possible
  to form the dimensionless combination $\lambda=H^2_\Lambda(G\hbar/c^3) \equiv  (L_P^2/L_\Lambda^2)$.
  Observations then require $(L_P^2/L_\Lambda^2) \lesssim 10^{-123}$ requiring enormous fine tuning. 
  
  In the early days, this was considered puzzling but 
most people believed that this number $\lambda$ is actually zero. The \cc\ problem in those days was to understand
why it is strictly zero. Usually, the vanishing of a   
constant (which could have appeared in the low energy sector of the theory) indicates an underlying symmetry of the 
theory. For example, the vanishing of the mass of the photon is closely related to
the gauge invariance of electromagnetism. No such symmetry principle is known to operate
at low energies which made this problem very puzzling. There is a symmetry --- called
supersymmetry --- which does ensure that $\lambda =0$ but it is known that supersymmetry is 
broken at sufficiently high energies and hence cannot explain the observed value of $\lambda$.

Given the observational evidence for  dark energy in the universe and the fact that
the simplest candidate for dark energy, consistent with all observations today, is a cosmological 
constant with $\lambda \approx 10^{-123}$ the cosmological constant problem has got linked to the 
problem of dark energy in the universe. So, if we accept  the simplest  interpretation of the
current observations, we need to explain why cosmological constant is non zero
and has this small value. It should, however, be stressed that these are 
logically independent issues. \textit{Even if all the observational evidence for dark energy goes
away we still have a problem --- viz., explaining why $\lambda$ is zero.}

There is another, related, aspect to \cc\ problem which need to be stressed. In conventional approach to gravity, one derives the equations of motion
from a Lagrangian $\mathcal{L}_{\rm tot} = \mathcal{L}_{\rm grav}(g) + \mathcal{L}_{\rm matt}(g,\phi)$ where
$\mathcal{L}_{\rm grav}$ is the gravitational Lagrangian dependent on the metric and its derivative
and $\mathcal{L}_{\rm matt}$ is the matter Lagrangian which depends on both the metric and the 
matter fields, symbolically denoted as $\phi$. In such an approach, the cosmological constant can be introduced via two different routes
which are conceptually different but operationally the same. 
First, one may decide
to take the gravitational Lagrangian to be $\mathcal{L}_{\rm grav} =(2\kappa)^{-1}(R-2\Lambda_g)$
where $\Lambda_g$ is a parameter in the  (low energy effective) action  just like
the Newtonian gravitational constant $\kappa$. 
The second route  is by
shifting  the matter Lagrangian by $\mathcal{L}_{\rm matt}\to \mathcal{L}_{\rm matt} - 2\lambda_m$. Such a shift is clearly
equivalent to adding a cosmological constant $2\kappa\lambda_m$ to the
$\mathcal{L}_{\rm grav}$. In general, what can be observed through gravitational interaction 
is the combination $\Lambda_{\rm tot} = \Lambda_g
+ 2\kappa\lambda_m$.  

It is now clear that there are two distinct aspects to the  cosmological
constant problem. The first question is why $\Lambda_{\rm tot} $ is very small
when expressed in natural units. Second, since $\Lambda_{\rm tot}$ could have
had two separate contributions from the gravitational and matter sectors, why
does the \textit{sum} remain so fine tuned? This question is particularly relevant because it is believed that our universe went through several phase transitions in the course of its  evolution, each of which shifts the energy momentum tensor of matter by $T^a_b\to T^a_b+L^{-4}\delta^a_b$ where $L$ is the scale characterizing the transition. For example, the GUT and Weak Interaction scales are about $L_{GUT}\approx 10^{-29}$ cm, $L_{SW}\approx 10^{-16}$ cm respectively which are tiny compared to $L_{\Lambda}$. 
Even if we take a more pragmatic approach, the observation of Casimir effect in the lab sets a bound that $L<\mathcal{O}(1)$ nanometer, leading to a $\rho$ which is about $10^{12}$ times the observed value \cite{gaurang}. 

Finally, I will comment on two other issues related to \cc\ which appear frequently in the literature.
The first one is what could be called the ``why now'' problem of the \cc. 
How come the energy density contributed by  the \cc\ (treated as the dark energy) is comparable to the energy density of the rest of the matter at the \textit{current epoch} of the universe? I  do not believe this is an \textit{independent} problem; 
if we have a viable theory predicting a particular numerical value for $\lambda$, then
the energy density due to this \cc\ will be comparable to the rest of the energy density at
\textit{some} epoch. So the real problem is in understanding the numerical value of $\lambda$;
once that problem is solved the `why now' issue will take care of itself. In fact, we do not have a viable
theory to predict the current energy densities of any component  which populates the universe, let alone the dark energy!.
For example, the energy density of radiation today is computed from its temperature which is an 
observed parameter --- there is no theory which tells us that this temperature has to be
2.73 K when, say,  galaxy formation has taken place for certain billion number of years.

One also notices in the   literature a discussion of the contribution of 
the zero point energies of the quantum fields to the \cc\ which is often  misleading, if not incorrect. What is usually done is to attribute a zero-point-energy $(1/2)\hbar\omega$ to each mode of the field and add up all these energies with an ultra violet cut-off. For an electromagnetic field, for example, this will lead to an integral proportional to
\begin{equation}
\rho_0=\int_0^{k_{max}} dk\ k^2 \hbar k\propto k_{max}^4
\end{equation}  
which will give $\rho_0\propto L_P^{-4}$ if we invoke a Planck scale cut-off with $k_{max}=L_P^{-1}$. It is then claimed that, this $\rho_0$ will contribute to the cosmological constant. There are several problems with such a naive analysis. First, the $\rho_0$ computed above can be easily eliminated by the normal ordering prescription in quantum field theory and what one really should compute is the fluctuations in the vacuum energy --- not the vacuum energy itself. Second, even if we take the nonzero value of $\rho_0$ seriously, it is not clear this has anything to do with a \cc. The energy momentum tensor due to the \cc\ has a very specific form $T^a_b\propto\delta^a_b$ and its trace is nonzero. The electromagnetic field, for example, has a stress tensor with zero trace, $T^a_a=0$; hence in the vacuum state
the expectation value of the trace, $\langle \mathrm{vac}|T^a_a|\mathrm{vac}\rangle$, will vanish, showing that
 the equation of state of the bulk electromagnetic vacuum is still
 $\rho_0=3p_0$ which does not lead to a \cc\ . (The trace anomaly will not work in the case of electromagnetic field.) So the naive calculation of vacuum energy density with a cutoff and the claim that it contributes to \cc\ is not an accurate statement in many cases. 
  
\section{What if dark energy is not the \cc\ ?} 

A nice possibility would be to postulate that $\lambda =0$ and come up with a symmetry principle
which will explain why this is the case. One probably has a greater chance of success in such 
an attempt than in coming up with an explanation for $\lambda \approx 10^{-123}$.
But then, one needs to provide an alternative explanation for the dark energy observations.
We shall now discuss two classes of such explanations, one which uses conventional physics and the other which is totally speculative --- and conclude that both are not viable!

\subsection{Conservative explanations of dark energy}

One of the \textit{least} esoteric ideas regarding the dark energy
is that the cosmological constant term in the  equations arises because we have not calculated the energy density driving the expansion of the universe correctly. This idea arises as follows:  The energy momentum tensor of the real universe, $T_{ab}(t,{\bf x})$ is inhomogeneous and anisotropic. If  we could solve the exact Einstein's equations
$G_{ab}[g]=\kappa T_{ab}$ with it as the source we will be led to a  complicated metric $g_{ab}$. 
The metric describing the large scale structure of the universe should be obtained by averaging this exact solution over a large enough scale, leading to $\langle g_{ab}\rangle $. But since we cannot solve  exact Einstein's equations, what we actually do is to average the stress tensor {\it first} to get $\langle T_{ab}\rangle $ and {\it then} solve Einstein's equations. But since $G_{ab}[g]$ is  nonlinear function of the metric, $\langle G_{ab}[g]\rangle \neq G_{ab}[\langle g\rangle ]$ and there is a discrepancy. This is most easily seen by writing
\begin{equation}
G_{ab}[\langle g\rangle ]=\kappa [\langle T_{ab}\rangle  + \kappa^{-1}(G_{ab}[\langle g\rangle ]-\langle G_{ab}[g]\rangle )]\equiv \kappa [\langle T_{ab}\rangle  + T_{ab}^{corr}]
\end{equation}
If --- based on observations --- we take the $\langle g_{ab}\rangle $ to be the standard Friedman metric, this equation shows that it has, as its  source,  \textit{two} terms:
The first is the standard average stress tensor and the second is a purely geometrical correction term
$T_{ab}^{corr}=\kappa^{-1}(G_{ab}[\langle g\rangle ]-\langle G_{ab}[g]\rangle )$ which arises because of nonlinearities in the Einstein's theory that  leads to $\langle G_{ab}[g]\rangle \neq G_{ab}[\langle g\rangle ]$. If this term can mimic the \cc\ at large scales there will be no need for dark energy and --- as a bonus --- one will solve the ``why now'' problem!
 
 To make this idea concrete, we have to identify an  effective expansion factor $a_{eff}(t)$
 of an inhomogeneous universe (after suitable averaging), and determine the equation of motion satisfied by it. The hope is that it will be sourced by terms so as to have $\ddot a_{eff}(t)>0$ while the standard matter (with $(\rho + 3p)>0$) leads to deceleration of standard expansion factor $a(t)$. Since any correct averaging of
 positive quantities  in $(\rho + 3p)$ will not lead to a negative quantity, the real hope is in defining $a_{eff}(t)$ and obtaining its dynamical equation such that
 $\ddot a_{eff}(t)>0$.  In spite of some recent attention this idea has received \cite{flucde} it is doubtful whether it will lead to the correct result when implemented properly. The reasons for my skepticism are the following:
 
 \begin{itemize}
 \item
 It is obvious that $T_{ab}^{corr}$ is --- mathematically speaking --- non-zero (for an explicit computation, in a completely different context of electromagnetic plane wave, see \cite{gofemw}); the real question  is how big is it compared to $T_{ab}$. It seems unlikely that when properly done, we will get a large effect for the simple reason that the amount of mass which is contained in the nonlinear regimes in the universe today is subdominant.

 \item
 Any calculation in linear theory or any calculation in which special symmetries are invoked will be inconclusive in settling this issue.
 The key question, of identifying a suitable analogue of expansion factor from an averaged geometry, is nontrivial and it is not clear that the answer will be unique. To illustrate this point by an extreme  example,
 suppose we decide to call $a(t)^n$ with, say $n>2$ as the effective expansion factor i.e.,  $a_{\rm eff}(t)=a(t)^n$; obviously $\ddot a_{\rm eff}$ can be positive (`accelerating universe') even with 
 $\ddot a$ being negative. So, unless one has a \textit{unique} procedure to identify the expansion factor of the average universe, it is difficult to settle the issue.
 
\item
 This approach is   strongly linked to explaining the acceleration as observed by SN. Even if we decide to completely ignore all SN data, we still have reasonable evidence for dark energy and it is not clear how this approach can tackle such evidence.
 
\end{itemize}
 
Another equally conservative explanation for the cosmic acceleration will be that we are located in a large underdense region in the universe; so that, locally, the underdensity acts like negative mass and produces a repulsive force. While there has been some discussion in the literature \cite{Hbubble} as to whether observations indicate such a local `Hubble bubble', this does not seem to be a tenable explanation that one can take seriously at this stage. Again, CMBR observations indicating dark energy, for example, will not be directly affected by this feature though one does need to take into account the effect of the local void.

\subsection{Dark Energy from scalar fields}

The most popular alternative to the \cc\  uses a scalar field $\phi$ with a suitably chosen potential $V(\phi)$ so as to make the vacuum energy vary with time. The hope then is that, one can find a model in which the current value can be explained naturally without any fine tuning. The scalar fields come in different shades and hues like quintessence, K-essence, tachyonic fields amongst others.
For
  a small sample of recent [$\gtrsim$ 2006]  papers, see \cite{phiindustry,tachyon}.
 
   Since  the quintessence field (or the tachyonic field)   has
   an undetermined free function $V(\phi)$, it is possible to choose this function
  in order to produce a given expansion history of the universe characterized by the function $H(a)=\dot a/a$ expressed in terms of $a$. 
  To see this explicitly, let
   us assume that the universe has two forms of energy density with $\rho(a) =\rho_{\rm known}
  (a) + \rho_\phi(a)$ where $\rho_{\rm known}(a)$ arises from all  known forms of source 
  (matter, radiation, ...) and
  $\rho_\phi(a) $ is due to a scalar field.  
  Let us first consider quintessence models  with the Lagrangian:
   \begin{equation}
  \mathcal{L}_{\rm quin} = \frac{1}{2} \partial_a \phi \partial^a \phi - V(\phi)
  \end{equation}
   Here,  the potential is given implicitly by the form
  \cite{tptachyon,ellis}
  \begin{equation}
  V(a) = \frac{1}{16\pi G} H (1-Q)\left[6H + 2aH' - \frac{aH Q'}{1-Q}\right]
  \label{voft}
   \end{equation} 
    \begin{equation}
    \phi (a) =  \left[ \frac{1}{8\pi G}\right]^{1/2} \int \frac{da}{a}
     \left[ aQ' - (1-Q)\frac{d \ln H^2}{d\ln a}\right]^{1/2}
    \label{phioft}
    \end{equation} 
   where $Q (a) \equiv [8\pi G \rho_{\rm known}(a) / 3H^2(a)]$ and prime denotes differentiation with respect to $a$.
   Given any
   $H(a),Q(a)$, these equations determine $V(a)$ and $\phi(a)$ and thus the potential $V(\phi)$. 
   \textit{Every quintessence model studied in the literature can be obtained from these equations.}
  
  Similar results exists for the tachyonic scalar field as well \cite{tptachyon} which has the Lagrangian:
  \begin{equation}
 \mathcal{L}_{\rm tach}
  = -V(\phi) [1-\partial_a\phi\partial^a\phi]^{1/2}
  \end{equation} 
 Given
  any $H(a)$, one can construct a tachyonic potential $V(\phi)$ so that the scalar field is the 
  source for the cosmology. The equations determining $V(\phi)$  are now given by:
  \begin{equation}
  \phi(a) = \int \frac{da}{aH} \left(\frac{aQ'}{3(1-Q)}
   -\frac{2}{3}\frac{a H'}{H}\right)^{1/2}
  \label{finalone}
  \end{equation}
   \begin{equation}
   V(a) = \frac{3H^2}{8\pi G}(1-Q) \left( 1 + \frac{2}{ 3}\frac{a H'}{ H}-\frac{aQ'}{3(1-Q)}\right)^{1/2}
   \label{finaltwo}
   \end{equation}
   Equations (\ref{finalone}) and (\ref{finaltwo}) completely solve the problem. Given any
   $H(a)$, these equations determine $V(a)$ and $\phi(a)$ and thus the potential $V(\phi)$. 
A wide variety of phenomenological models with time dependent
  \cc\ have been considered in the literature; all of these can be 
   mapped to a 
  scalar field model with a suitable $V(\phi)$.

  It is very doubtful whether this ---  rather popular ---  approach, based on scalar fields, has helped us to understand the nature of the dark energy
  at any deeper level. These
  models, viewed objectively, suffer from several shortcomings:
   
 \begin{itemize}
  \item
  The most serious problem with them is that they have no predictive power. As explicitly demonstrated above, virtually every form of $a(t)$ can be modeled by a suitable ``designer" $V(\phi)$.
  \item
  We see from the above discussion that even when $w(a)$ is determined by observations, it is not possible to proceed further and determine
  the nature of the scalar field Lagrangian. The explicit examples given above show that there
  are {\em at least} two different forms of scalar field Lagrangians --- corresponding to
  the quintessence or the tachyonic field --- which could lead to
  the same $w(a)$. (See the first paper in ref.\cite{tptirthsn1} for an explicit example of such a construction.)
  
  \item
  By and large, the potentials  used in the literature have no natural field theoretical justification. All of them are non-renormalisable in the conventional sense and have to be interpreted as a low energy effective potential in an ad hoc manner.
  \item
  One key difference between \cc\ and scalar field models is that the latter lead to a $(p/\rho) \equiv w(a)$ which varies with time. So they are worth considering if the  observations have suggested a varying $w$, or if observations have ruled out $w=-1$ at the present epoch. However, all available observations are consistent with \cc\ ($w=-1$) and --- in fact --- the possible variation of $w$ is strongly constrained \cite{jbp}. 
 \end{itemize}

  As an aside, let us note that in drawing conclusions from the  observational data, one should be careful about the hidden assumptions in the statistical analysis. Claims regarding  $w$ depends crucially on the data sets used, priors which are assumed and possible parameterizations which are adopted. (For more details related to these issues, see the last reference in \cite{jbp}.) It is fair to say that all currently available data is consistent with $w=-1$. Further, there is some amount of tension between WMAP and SN-Gold data with the recent SNLS data \cite{snls} being more concordant with WMAP than the SN Gold data.

One also needs to remember that, 
 for the scalar field models to work,
we first need to find a mechanism which will make the \cc\ vanish. In other words,  all the scalar field potentials require fine tuning of the parameters in order to be viable. This is obvious in the quintessence models in which adding a constant to the potential is the same as invoking a \cc. If we shift $\mathcal{L}\to \mathcal{L}_{\rm matt} - 2\lambda_m$ in an otherwise successful scalar field model for dark energy, we end up `switching on' the cosmological constant and raising the problems again.

\section{Cosmological Constant as dark energy}

Even if all the evidence for dark energy disappears within a decade, we still need to understand why \cc\ is zero and much of what I have to say in the sequel will remain relevant. I stress this because there is a recent tendency to forget the fact that the problem of the \cc\ existed (and was recognized as a problem) long before the observational evidence for dark energy, accelerating universe etc cropped up. In this sense, \cc\ problem has an important theoretical dimension which is distinct from what has been introduced by the observational evidence for dark energy.   

Though invoking the \cc\ as the candidate for dark energy leads to well known problems mentioned earlier, it is also the most economical (just one number) explanation for all the observations. 
Therefore it is worth examining this idea in detail and ask how these problems can be tackled.

 If the \cc\ is nonzero, then classical gravity will be described by the three constants $G,c$ and $\Lambda\equiv L_\Lambda^{-2}$. Since $\Lambda(G\hbar/c^3)\equiv (L_P/L_\Lambda)^2\approx 10^{-123}$, it is obvious that the \cc\ is actually telling us something regarding \textit{quantum gravity}, indicated by the combination $G\hbar$. \textit{An acid test for any quantum gravity model will be its ability to explain this value;} needless to say, all the currently available models --- strings, loops etc.  --- flunk this test.

While the occurrence of $\hbar$ in $\Lambda(G\hbar/c^3)$ shows that it is a relic of a quantum gravitational effect (or principle) of unknown nature, 
 \cc\ problem is an infrared problem \textit{par excellence} in terms of the energy scales which are involved.
 This is a somewhat unusual possibility of  a high energy phenomenon leaving a low energy relic and an analogy will be helpful to illustrate this idea \cite{choices}. Suppose we solve the Schrodinger equation for the Helium atom for the quantum states of the two electrons $\psi(x_1,x_2)$. When the result is compared with observations, we will find that only half the states --- those in which  $\psi(x_1,x_2)$ is antisymmetric under $x_1\longleftrightarrow x_2$ interchange --- are realized in nature. But the low energy Hamiltonian for electrons in the Helium atom has no information about
this effect! Here is a low energy (IR) effect which is a relic of relativistic quantum field theory (spin-statistics theorem) that is  totally non perturbative, in the sense that writing corrections to the Hamiltonian of the Helium atom  in some $(1/c)$ expansion will {\it not} reproduce this result. I suspect the current value of \cc\ is related to quantum gravity in a similar spirit. There must exist a deep principle in quantum gravity which leaves its non-perturbative trace even in the low energy limit
that appears as the \cc.
We shall now attempt a more quantitative discussion of these possibilities.

 \subsection{Area scaling law for energy fluctuations}\label{sec:ccnatural}

  Given the  theory with two length scales $L_P$ and $L_\Lambda$, one can construct two energy scales
 $\rho_{_{\rm UV}}=1/L_P^4$ and $\rho_{_{\rm IR}}=1/L_\Lambda^4$ in natural units ($c=\hbar=1$). There is sufficient amount of justification from different theoretical perspectives
 to treat $L_P$ as the zero point length of spacetime \cite{zeropoint}, giving a natural interpretation to $\rho_{_{\rm UV}}$. The second one, $\rho_{_{\rm IR}}$ also has a natural interpretation. Since the universe  dominated by a \cc\ at late times will be  asymptotically DeSitter with $a(t)\propto \exp (t/L_\Lambda) $ at late times, it will have a horizon and associated thermodynamics \cite{ghds} with a  temperature
 $T=H_\Lambda/2\pi$. The corresponding thermal energy density is $\rho_{thermal}\propto T^4\propto 1/L_\Lambda^4=
 \rho_{_{\rm IR}}$. Thus $L_P$ determines the \textit{highest} possible energy density in the universe while $L_\Lambda$
 determines the {\it lowest} possible energy density in this universe. As the energy density of normal matter drops below this value, $\rho_{IR}$, the thermal ambiance of the DeSitter phase will remain constant and provide the irreducible `vacuum noise'. The observed dark energy density is the the geometric mean 
\begin{equation}
 \rho_{_{\rm DE}}=\sqrt{\rho_{_{\rm IR}}\rho_{_{\rm UV}}}=\frac{1}{L_P^2L_\Lambda^2}
 \label{geomean}
\end{equation} 
 of these two energy densities. If we define a dark energy length scale $L_{DE}$  such that $\rho_{_{\rm DE}}=1/L_{DE}^4$ then $L_{DE}=\sqrt{L_PL_\Lambda}$ is the geometric mean of the two length scales in the universe. 
 
It is possible to interpret  this relation  along the following lines: Consider a 3-dimensional region of size $L$ with a bounding area which scales as $L^2$. Let us assume that we
 associate with
 this region  $N$ microscopic cells of size $L_P$ 
each having a Poissonian fluctuation in energy of amount $E_P\approx 1/L_P$. Then the mean square fluctuation of energy in this region will be $(\Delta E)^2\approx NL_P^{-2}$ corresponding to the energy density
$\rho=\Delta E/L^3=\sqrt{N}/L_PL^3$. If we make the usual assumption that $N=N_{vol}\approx (L/L_P)^3$, this will give
\begin{equation}
\rho=\frac{\sqrt{N_{vol}}}{L_PL^3}=\frac{1}{L_P^4}\frab{L_P}{L}^{3/2} \quad \text {(bulk\ fluctuations)}
\end{equation} 
On the other hand, if we assume that (for reasons which are unknown), the relevant degrees of freedom scale as the surface area of the region, then $N=N_{sur}\approx (L/L_P)^2$
and the relevant energy density is
\begin{equation}
\rho=\frac{\sqrt{N_{sur}}}{L_PL^3}=\frac{1}{L_P^4}\frab{L_P}{L}^2=\frac{1}{L_P^2L^2} \quad \text {(surface\ fluctuations)}
\label{sur}
\end{equation}
If we take $L\approx L_\Lambda$, the surface fluctuations in \eq{sur} give precisely the geometric mean in \eq{geomean} which is observed. On the other hand, the bulk \textit{fluctuations} lead to an energy density which is larger by a factor 
$(L/L_P)^{1/2}$. 
 Of course, if we do not take fluctuations in energy but coherently add them, we will get $N/L_PL^3$ which is $1/L_P^4$ for the bulk and $(1/L_P)^4(L_P/L)$
for the surface. In summary, we have the following hierarchy:
\begin{equation}
\rho=\frac{1}{L_P^4}\times \left[1,\frab{L_P}{L},
\frab{L_P}{L}^{3/2},
\frab{L_P}{L}^2,
\frab{L_P}{L}^4 .....\right]
\end{equation} 
in which the first one arises by coherently adding energies $(1/L_P)$ per cell with
$N_{vol}=(L/L_P)^3$ cells; the second arises from coherently adding energies $(1/L_P)$ per cell with
$N_{sur}=(L/L_P)^2$ cells; the third one is obtained by taking \textit{fluctuations} in energy and using $N_{vol}$ cells; the fourth from energy fluctuations with $N_{sur}$ cells; and finally the last one is the thermal energy of the DeSitter space if we take $L\approx L_\Lambda$;  clearly the further terms are irrelevant due to this vacuum noise. 

Of all these, the only viable possibility is what arises if we assume that: 
(a)
The number of active degrees of freedom in a region of size $L$ scales as $N_{sur}=(L/L_P)^2$.
(b)
It is the \textit{fluctuations} in the energy that contributes to the cosmological constant \cite{cc1,cc2} and the bulk energy does not gravitate.

It has been demonstrated recently
\cite{tpholo}   that it is possible to obtain 
classical relativity from purely thermodynamic considerations in which the surface term of the gravitational action plays a crucial role.  
The area scaling is familiar from the usual result that entropy of horizons scale as area. 
(Further, in cases like Schwarzschild black hole, one cannot even properly define the volume inside a horizon.)
In fact, one can argue from general considerations that the entropy associated with \textit{any} null surface should be $(1/4)$ per unit area and will be observer dependent.  A null surface, obtained as a limit of a sequence of timelike surfaces (like the $r=2M$ obtained from $r=2M+k$ surfaces with $k\to 0^+$ in the case of the Schwarzschild black hole), `loses' one dimension in the process (e.g., $r=2M+k$ is 3-dimensional and timelike for $k>0$ but is 2-dimensional and null for $k=0$) suggesting that the scaling of degrees of freedom has to change appropriately.
It is difficult to imagine that these features are unconnected and accidental and we will discuss these ideas further in the next Section.

 \section{An alternative perspective: Emergent Gravity}
 
I will now describe an alternative perspective   in which gravity is treated as an emergent phenomenon -- like elasticity -- and argue that such a perspective is indeed \textit{necessary} to succeed in solving the \cc\ problem. To do this, I will first identify the key ingredient of the \cc\ problem and try to address it head on.

 \subsection{Why do we need a new perspective on gravity?}
 
 The equations of motion of gravity is obtained in the  conventional approach to gravity
from a Lagrangian $\mathcal{L}_{\rm tot} = \mathcal{L}_{\rm grav}(g) + \mathcal{L}_{\rm matt}(g,\phi)$ where
$\mathcal{L}_{\rm grav}$ is the gravitational Lagrangian dependent on the metric and its derivative
and $\mathcal{L}_{\rm matt}$ is the matter Lagrangian which depends on both the metric and the 
matter fields, symbolically denoted as $\phi$. This total Lagrangian is integrated
over the spacetime volume with the covariant measure $\sqrt{-g} d^4x$ to obtain the 
action. 

Suppose we now add a constant $(- 2\lambda_m)$ to the matter Lagrangian thereby inducing the change
 $\mathcal{L}_{\rm matt}\to \mathcal{L}_{\rm matt} - 2\lambda_m$. The equations
of motion for matter are invariant under such a transformation which implies that --- in the 
absence of gravity --- we cannot determine the value of $\lambda_m$. 
The transformation $\mathcal{L}\to \mathcal{L}_{\rm matt} - 2\lambda_m$  is a symmetry
of the matter sector (at least at scales below the scale of supersymmetry breaking; we shall ignore supersymmetry in what follows). But,
in the conventional approach, gravity breaks this
symmetry.  \textit{This is the root cause of the  cosmological constant problem.}
As long as
gravitational field equations are of the form $E_{ab} = \kappa T_{ab}$ where $E_{ab}$ is some geometrical quantity (which is $G_{ab}$ in Einstein's theory) the theory
cannot be invariant under the shifts of the form $T^a_b \to T^a_b +\rho \delta^a_b$.
Since such shifts are allowed by the matter sector, it is very difficult to imagine a definitive  solution
to cosmological constant problem within the conventional approach to gravity.

If metric represents the gravitational degree of freedom that is varied in the action and we demand full general covariance,  we cannot  avoid $\mathcal{L}_{matter}\sqrt{-g}$ coupling and cannot obtain  of the equations of motion which are invariant under the shift
$T_{ab}\to T_{ab}+\Lambda g_{ab}$.
 Clearly a new, drastically different, approach to gravity is required.
 We need to look for an approach which has the following ingredients
\cite{gr06}:
 
 To begin with, the field equations must remain invariant under the shift $\mathcal{L}_{matt}\to \mathcal{L}_{matt}+\lambda_m$ of the matter Lagrangian $\mathcal{L}_{matt}$ by a constant $\lambda_m$. That is, we need to have some kind of `gauge freedom' to absorb any $\lambda_m$. 
General covariance requires using the integration measure $\sqrt{-g}d^Dx$ in actions. Since we do not want to restrict general covariance but at the same time do not want this coupling to metric tensor via $\sqrt{-g}$, it follows that \textit{Metric cannot be the dynamical variable in our theory.}
Secondly, even if we manage to obtain a theory in which gravitational action is invariant under the shift $T_{ab}\to T_{ab}+\Lambda g_{ab}$,  we would have only   succeeded in making gravity  decouple from the bulk vacuum energy. While this is considerable progress, there still remains the second issue of explaining the observed value of the \cc. Once the bulk value of the \cc\ (or vacuum energy) decouples from gravity, \textit{classical} gravity becomes immune to \cc; that is, the bulk classical \cc\ can be gauged away.
Any observed value of the \cc\ has to be necessarily a \textit{quantum} phenomenon arising as a relic of microscopic spacetime fluctuations. 
The discussion in section \ref{sec:ccnatural}, especially 
 Eq.~(\ref{sur}), shows that the
 relevant degrees of freedom should be linked to  surfaces in spacetime rather than
bulk regions.  The observed \cc\ is  a relic of quantum gravitational physics and should arise from  degrees of freedom which scale as the surface area. 

Thus, in an approach in which the surface degrees of freedom play the dominant role, rather than bulk degrees of freedom, we have a hope for obtaining the correct value for the \cc.
One should then  obtain a theory of gravity which is more general than Einstein's theory with the latter emerging as a low energy approximation.

\subsection{Micro-structure  of the spacetime}

For reasons described above, we abandon the usual picture of treating the
metric as  the fundamental dynamical degrees of freedom of the theory and treat it as
 providing a
coarse grained description of the spacetime at macroscopic scales,
somewhat like the density of a solid ---  which has no meaning at atomic  
scales \cite{elastic}.  The unknown, microscopic degrees of freedom of
spacetime (which should be analogous to the atoms in the case of
solids), will play a role only when spacetime is probed at Planck
scales (which would be analogous to the lattice spacing of a solid
\cite{zeropoint}).  
 
 Some further key insight can be obtained by noticing that 
 in the study of ordinary solids, one can distinguish between three levels of description. At the macroscopic level,
we have the theory of elasticity which has a life of its own and can be developed purely phenomenologically.  At the other extreme,  the microscopic description of a solid will be in terms of the statistical mechanics of a lattice of atoms and their interaction.

Both of these are well known; but interpolating between these two limits  is the thermodynamic description of a solid at finite temperature \textit{which provides a crucial window into the existence of the corpuscular substructure of solids.} As Boltzmann told us, heat is a form of motion
and we will not have the thermodynamic layer of description if  matter is a continuum all the way to the finest scales and atoms did not exist! \textit{The mere existence of a thermodynamic layer in the description is proof enough that there are microscopic degrees of freedom. } 

 The situation is similar in the case of 
 the spacetime \cite{roorkee}. Again we should have three levels of description. The macroscopic level is the smooth spacetime continuum with a metric tensor $g_{ab}(x^i)$ and the  equations governing the metric have the same status as the phenomenological equations of elasticity. At the microscopic level, we  expect a quantum description in terms of the `atoms of spacetime' and some associated degrees of freedom $q_A$ which are still elusive. But what is crucial is the existence of an interpolating layer of thermal phenomenon associated with null surfaces in the spacetime.  Just as a solid  cannot exhibit thermal phenomenon if it does not have microstructure,  \textit{thermal nature of horizon, for example, cannot arise without the spacetime having a microstructure. } 

In such a picture, we normally expect the  microscopic structure of spacetime
to manifest itself only at Planck scales or near singularities of the
classical theory. However, in a manner which is not fully understood,
the horizons ---  which block information from certain classes of
observers --- link \cite{magglass} certain aspects of microscopic
physics with the bulk dynamics, just as thermodynamics can provide a
link between statistical mechanics and (zero temperature) dynamics of
a solid. The  reason is probably related to the fact that horizons
lead to infinite redshift, which probes \textit{virtual} high energy
processes.

The following three results, showing a 
  fundamental relationship between the dynamics of gravity and thermodynamics of horizons \cite{paddy1}
  strongly support  the above point of view:

\begin{itemize}
\item
The dynamical equations governing the metric
 can be interpreted as a thermodynamic relation closely related to
 the thermodynamics of horizons. An explicit example  was provided in ref. \cite{paddy2}, in the case of spherically symmetric horizons in four
dimensions in which it was shown that, Einstein's equations can be interpreted as a 
thermodynamic relation $TdS=dE+PdV$ arising out of virtual
radial displacements of the horizon. Further work showed that this result is valid in \textit{all} the cases for which explicit computation can be carried out --- like in the
Friedmann models 
\cite{rongencai} as well as for rotating and time dependent horizons
in Einstein's theory \cite{dawood-sudipta-tp}. 
\item
The standard Lagrangian in Einstein's theory has the structure $\mathcal{L}_{EH}\propto R\sim (\partial g)^2+ {\partial^2g}$.
In the usual approach the surface term arising from  $\mathcal{L}_{sur}\propto \partial^2g$ has to be ignored or canceled to get Einstein's equations from  $\mathcal{L}_{bulk}\propto (\partial g)^2$.  
But there is a 
peculiar (unexplained) relationship \cite{tpholo} between $\mathcal{L}_{bulk}$ and $\mathcal{L}_{sur}$:
\begin{equation}
    \sqrt{-g}\mathcal{L}_{sur}=-\partial_a\left(g_{ij}
\frac{\partial \sqrt{-g}\mathcal{L}_{bulk}}{\partial(\partial_ag_{ij})}\right)
\end{equation}
This shows that the gravitational action is `holographic' with the same information being coded in both the bulk and surface terms and one of them should be sufficient. 
\item
One can indeed obtain Einstein's equations from an action principle which  uses \textit{only} the surface term and the  virtual displacements of horizons \cite{paris,gr06}. 
It is possible to determine the form of this surface term from general considerations. If we now demand that the action should not receive contributions for radial displacements of the horizons, defined in a particular manner using local Rindler horizons, one can obtain --- at the lowest order --- the equations
 \begin{equation}
(G_{ab}-\kappa T_{ab})\xi^a\xi^b=0
\label{nullvec}
\end{equation} 
where $\xi^a$ is a null vector.  Demanding the validity of \eqn{nullvec} in all local Rindler frames then leads to Einstein's theory with the cosmological constant emerging as an integration constant. Note that \eqn{nullvec} is invariant under the constant shift of matter Lagrangian, making gravity immune to bulk cosmological constant.
Since the surface term has the thermodynamic interpretation as the entropy of horizons, this establishes a direct connection between spacetime dynamics and horizon thermodynamics.
\item
Further work has shown that \textit{all the above results extend  beyond Einstein's theory.}
 The connection between field equations and the thermodynamic relation $TdS=dE+PdV$ 
 is not restricted to
Einstein's theory  alone, but is in fact true for the
case of the generalized, higher derivative \LL gravitational theory in
$D$ dimensions as well \cite{aseem-sudipta, cai2}. The same is true \cite{ayan} for the holographic structure of the action functional: the \LL action has the same structure and --- again --- the entropy of the horizons is related to the surface term of the action. \textit{These results show that the thermodynamic description is far more general than just Einstein's theory} and occurs in a wide class of theories in which the metric determines the structure of the light cones and null surfaces exist blocking the information.
\end{itemize}

The conventional approach to gravity fails to provide any clue 
 regarding the thermodynamic aspects of gravity
 just as Newtonian continuum mechanics --- without corpuscular, discrete, substructure for matter --- cannot explain thermodynamic phenomena.
A natural explanation for these results requires a different approach to spacetime dynamics which I will now outline. (More details can be found in ref. \cite{grgreview} )

\subsection{Surfaces in spacetime: key to the new paradigm}

In obtaining 
 the relation between gravitational dynamics and horizon thermodynamics, 
 one  treats the null surfaces (which act as horizons) as the limit of a sequence of, say, timelike surfaces. The virtual displacements of the horizon  in the direction normal to the surfaces
 will be
  used in the action principle.
  All these suggest that one may be able to obtain a more formal description of the theory in terms of deformation of surfaces in spacetime. I will now describe one such model which is unreasonably successful.

To set the stage, let us suppose there are certain microscopic --- as yet unknown --- degrees of freedom $q_A$, analogous to the atoms in the case of solids, described by some microscopic action functional $A_{micro}[q_A]$. In the case of a solid, the relevant long-wavelength elastic dynamics is captured by  the \textit{displacement vector field}
which occurs in the equation $x^a\to x^a+\xi^a(x)$ 
which is only very indirectly connected with the microscopic degrees of freedom. Similarly, 
in the case of spacetime,  we  need to introduce some other degrees of freedom, analogous to $\xi^a$ in the case of elasticity, and an effective action functional based on it.  
(As explained above, we do not want to use the metric as a dynamical variable.)
Normally, varying an action functional with respect to certain degrees of freedom will lead to equations of motion determining \textit{those} degrees of freedom.  But we now make an unusual demand that varying our  action principle with respect to some (non-metric) degrees of freedom should lead to an equation of motion \textit{determining the background metric} which remains non-dynamical.

Based on the role expected to be played by surfaces in spacetime, we shall take the relevant degrees of freedom to be the normalized vector fields $n_i(x)$ in the spacetime \cite{grgreview} with a  norm that  is fixed at every event but might vary from event to event: (i.e., $n_in^i\equiv\epsilon(x)$ with $\epsilon(x)$ being a fixed function which takes the values $0,\pm1$ at each event.) 
Just as the displacement vector $\xi^a$ captures the macro-description in case of solids, the  normalized vectors (e.g., local normals to surfaces) capture the essential macro-description in case of gravity in terms of an effective action $S[n^a]$. More formally, we expect the coarse graining of microscopic degrees of freedom to lead to an effective action in the  long wavelength limit:
\begin{equation}
\sum_{q_A}\exp (-A_{micro}[q_A])\longrightarrow \exp(-S[n^a])
\label{microtomac}
\end{equation} 
To proceed further we need to determine the nature of $S[n^a]$. The general form of $S[n^a]$ in such an effective description, at the quadratic order, will be:
\begin{equation}
S[n^a]=\int_\Cal{V}{d^Dx\sqrt{-g}}
    \left(4P_{ab}^{\ph{a}\ph{b}cd} \D_cn^a\D_dn^b - 
    T_{ab}n^an^b\right) \,,
\label{ent-func-2}
\end{equation}
where $P_{ab}^{\ph{a}\ph{b}cd}$ and $T_{ab}$ are two tensors and the signs, notation etc.  are chosen with hindsight. (We will see later that $T_{ab}$ can be identified with the matter stress-tensor.)
The full action for gravity plus matter will be taken to be $S_{tot}=S[n^a]+S_{matt}$ with:
\begin{equation}
S_{tot}=\int_\Cal{V}{d^Dx\sqrt{-g}}
    \left(4P_{ab}^{\ph{a}\ph{b}cd} \D_cn^a\D_dn^b - 
    T_{ab}n^an^b\right)+\int_\Cal{V}{d^Dx\sqrt{-g}} \mathcal{L}_{matt}
    \label{stotal}
\end{equation}  
with an important extra prescription: Since the gravitational sector is related to spacetime microstructure, we must \textit{first} vary the $n^a$ and \textit{then} vary the matter degrees of freedom. In the language of path integrals, we should  integrate out the gravitational degrees of freedom $n^a$ first and use the resulting action for the matter sector. 
 
We next address one crucial  difference 
between the dynamics in gravity and say, elasticity,  which we mentioned earlier. In the case of solids, one will write a similar functional for thermodynamic potentials in terms of the displacement vector $\xi^a$ and extremising it will
lead to an equation  \textit{which determines}
$\xi^a$. In the case of spacetime, we expect the variational principle to hold for   all vectors $n^a$ with a fixed norm and lead to a condition on  the \textit{background
metric.} Obviously, the action functional in \eqn{ent-func-2} must be rather special to accomplish this and one need to impose two restrictions on the coefficients $P_{ab}^{\ph{a}\ph{b}cd}$ and $T_{ab}$ to achieve this.  First, the tensor $P_{abcd}$ should
have the algebraic symmetries similar to the Riemann tensor $R_{abcd}$
of the $D$-dimensional spacetime. Second, we need:
\begin{equation}
\D_{a}P^{abcd}=0=\D_{a}T^{ab}\,.
\label{ent-func-1}
\end{equation}
In a
complete theory, the explicit form of $P^{abcd}$ will be determined by the
long wavelength limit of the microscopic theory just as the elastic
constants can --- in principle --- be determined from the microscopic
theory of the lattice. In the absence of such a theory, we can take a cue from
the renormalization group theory and expand $P^{abcd}$
in powers of  derivatives of
the metric \cite{paris,grgreview}. That is, we expect,
\begin{equation}
P^{abcd} (g_{ij},R_{ijkl}) = c_1\,\overset{(1)}{P}{}^{abcd} (g_{ij}) +
c_2\, \overset{(2)}{P}{}^{abcd} (g_{ij},R_{ijkl})  
+ \cdots \,,
\label{derexp}
\end{equation} 
where $c_1, c_2, \cdots$ are coupling constants and the successive terms progressively probe smaller and smaller scales.  The lowest order
term must clearly depend only on the metric with no derivatives. The next
term depends (in addition to metric) linearly on curvature tensor and the next one will be quadratic in curvature etc. It can be shown that  the m-th order term
which satisfies our constraints is \textit{unique} and is given by
\begin{equation}
\overset{(m)}{P}{}_{ab}^{\ph{a}\ph{b}cd}\propto
\AltC{c}{d}{a_3}{a_{2m}}{a}{b}{b_3}{b_{2m}}
\Riem{b_3}{b_4}{a_3}{a_3} \cdots
\Riem{b_{2m-1}}{b_{2m}}{a_{2m-1}}{a_{2m}} 
 =
\frac{\partial\LDm}{\partial R^{ab}_{{\ph{ab}cd}}}\,. 
\label{LL03}
\end{equation}
where $\AltC{c}{d}{a_3}{a_{2m}}{a}{b}{b_3}{b_{2m}}$ is the alternating tensor and the last equality shows that it
can be  expressed
as a derivative of the m th order \LL Lagrangian \cite{paris,grgreview,lovelock}, given by
\begin{equation}
\Cal{L}^{(D)} = \sD{c_m\LDm}\,~;~\Cal{L}^{(D)}_m = \frac{1}{16\pi}
2^{-m} \Alt{a_1}{a_2}{a_{2m}}{b_1}{b_2}{b_{2m}}
\Riem{b_1}{b_2}{a_1}{a_2} \Riem{b_{2m-1}}{b_{2m}}{a_{2m-1}}{a_{2m}}
\,,  
\label{LL222}
\end{equation}
where the $c_m$ are arbitrary constants and \LDm\ is the $m$-th
order \LL term and we assume
$D\geq2K+1$.
The lowest order term (which leads to Einstein's theory) is
\begin{equation}
\overset{(1)}{P}{}^{ab}_{cd}=\frac{1}{16\pi}
\frac{1}{2} \delta^{a_1a_2}_{b_1b_2} =\frac{1}{32\pi}
(\delta^a_c \delta^b_d-\delta^a_d \delta^b_c)
  \,.
\label{pforeh}
\end{equation} 
while the first order term  gives the  Gauss-Bonnet correction.
All higher orders terms are obtained in a similar manner.

In our paradigm based on \eqn{microtomac}, the field equations for gravity arise from extremising $S$ with respect to
variations of the  vector field $n^a$, with the constraint $\delta (n_an^a)=0$, and demanding that the
resulting condition holds for \textit{all normalized vector fields}.  
One can show \cite{paris,grgreview} that this leads to the field equations
\begin{equation}
16\pi\left[ P_{b}^{\ph{b}ijk}R^{a}_{\ph{a}ijk}-\frac{1}{2}\delta^a_b\LDm\right]=
 8\pi T{}_b^a +\Lambda\delta^a_b   
\label{ent-func-71}
\end{equation}
where $\Lambda$ is an undetermined integration constant.
These are identical to the field equations for the  \LL gravity with a cosmological constant arising as an undetermined integration constant.  To the lowest order, when we use \eqn{pforeh} for $P_{b}^{\ph{b}ijk}$, the \eqn{ent-func-71} reproduces Einstein's theory. More generally, we get Einstein's equations with
higher order corrections which are to be interpreted as emerging  
from the derivative expansion of the action functional as we probe smaller and smaller scales. 
Remarkably enough, we can derive not only Einstein's theory but even \LL theory from a dual description in terms on the normalized vectors in spacetime, \textit{without varying $g_{ab}$ in an action functional!}

The crucial feature of the coupling between matter and gravity through $T_{ab}n^an^b$ in \eq{stotal}
is that, under the shift $T_{ab}\to T_{ab}+\rho_0g_{ab}$, the $\rho_0$  term in the action in \eqn{ent-func-2} decouples from $n^a$ and becomes irrelevant:
\begin{equation}
\int_\Cal{V}{d^Dx\sqrt{-g}}T_{ab}n^an^b \to 
\int_\Cal{V}{d^Dx\sqrt{-g}} T_{ab}n^an^b +
\int_\Cal{V}{d^Dx\sqrt{-g}}\epsilon\rho_0
\end{equation} 
Since $\epsilon=n_an^a$ is not varied when $n_a$ is varied there is no coupling between $\rho_0$ and the dynamical variables $n_a$ and the theory is invariant under the shift  $T_{ab}\to T_{ab}+\rho_0g_{ab}$.
 We see that the condition $n_an^a=$ constant on the  dynamical variables have led to a `gauge freedom' which allows an arbitrary integration constant to appear in the theory which can absorb the bulk cosmological constant. 
   
To gain a bit more insight into what is going on, let us consider the on-shell value of the action functional in \eq{stotal}.  It can be shown that
 the \textit{on-shell} value is given by a  surface term 
which will lead to the entropy of the horizons (which will be 1/4 per unit transverse area in the case of general relativity).   Even in the case of  a theory with a  general $P^{ab}_{cd}$ it can be shown that the on-shell value of the action reduces to \cite{grgreview} the entropy of the horizons. The general expression is:
\begin{equation}
S|_{\Cal{H}} = \sD{4\pi m c_m \int_{\Cal{H}}{d^{D-2}x_{\perp} 
  \sqrt{\sigma}\Cal{L}^{(D-2)}_{(m-1)}}} 
 =\frac{1}{4}[{\rm Area}]_\perp +{\rm corrections}     
\label{ent-limit-2}
\end{equation} 
where $x_{\perp}$ denotes the transverse coordinates on the horizon \Cal{H},
$\sigma$ is the determinant of the intrinsic metric on \Cal{H} and we
have restored a summation over $m$ thereby giving the result for the
most general \LL case obtained as a sum of individual \LL Lagrangian's.  The expression in \eqn{ent-limit-2} \emph{is
  precisely the entropy of a general Killing horizon in \LL gravity}
based on the general prescription given by Wald 
and computed by several authors.  

This result shows that, in the semiclassical limit, in which the action can possibly be related to entropy, we reproduce the conventional entropy which scales as the area in Einstein's theory. Since the entropy counts the relevant degrees of freedom, this shows that the degrees of freedom which survives and contributes in the long wave length limit  scales as the area. The quantum fluctuations in these degrees of freedom can then lead to the correct, observed, value of the \cc. We will discuss this aspect briefly in the next section.

Our action principle is somewhat peculiar compared to the usual action principles in the sense that we have varied $n_a$ and demanded that the resulting equations hold for \textit{all} vector fields of constant norm. Our action principle actually stands for an infinite number of action principles, one for each vector field of constant norm! This class of \textit{all} $n^i$ allows an effective, coarse grained, description of some (unknown) aspects of spacetime micro physics. This is why we need to first vary $n_a$, obtain the equations constraining the background metric and then use the reduced action  to obtain the equations of motion for matter.  Of course, in most contexts, $\nabla_a T^a_b=0$ will take care of the dynamical equations for matter and these issues are irrelevant.

At this stage, it is not possible to proceed further and relate $n^i$ to some microscopic degrees of freedom $q^A$. This issue is conceptually similar to asking one to identify the atomic degrees of freedom, given the description of an elastic solid in terms of a displacement field $\xi^a$ --- which we know is virtually impossible. However, the same analogy tells us that the relevant degree of freedom in the long wavelength limit (viz. $\xi^a$ or $n^i$) can be completely different from the microscopic degrees of freedom and it is best to proceed phenomenologically.

 \subsection{Gravity as detector of the vacuum fluctuations}

The description of gravity  given above provides a natural back drop for gauging away the bulk value of the cosmological constant since it decouples from the dynamical degrees of freedom in the theory.  Once the bulk term is eliminated, 
what is observable through gravitational effects, in the correct theory of quantum gravity, should be the \textit{fluctuations} in the vacuum energy.
These fluctuations will be non-zero if the universe has a DeSitter horizon which provides a confining 
volume. In this paradigm the
 vacuum structure can readjust  to gauge away the bulk energy density $\rho_{_{\rm UV}}\simeq L_P^{-4}$ while quantum \textit{fluctuations} can generate
the observed value $\rho_{\rm DE}$. 

The role of energy fluctuations contributing to gravity also arises, more formally, when we study the question of \emph{detecting} the energy
density using gravitational field as a probe.
 Recall that a detector with a linear  coupling  to the {\it field} $\phi$
actually responds to $\langle 0|\phi(x)\phi(y)|0\rangle$ rather than to the field itself \cite{probe}. Similarly, one can use the gravitational field as a natural ``detector" of energy momentum tensor $T_{ab}$ with the standard coupling $L=\kappa h_{ab}T^{ab}$. Such a model was analyzed in detail in ref.~\cite{tptptmunu} and it was shown that the gravitational field responds to the two point function $\langle 0|T_{ab}(x)T_{cd}(y)|0\rangle $. In fact, it is essentially this fluctuations in the energy density which is computed in the inflationary models \cite{inflation} as the  {\it source} for gravitational field, as stressed in
ref.~\cite{tplp}. All these suggest treating the energy fluctuations as the physical quantity ``detected" by gravity, when
one  incorporates quantum effects.  

Quantum theory, especially the paradigm of renormalization group has taught us that the  concept of the vacuum
state  depends on the scale at which it is probed. The vacuum state which we use to study the
lattice vibrations in a solid, say, is not the same as vacuum state of the QED
 and it is not appropriate to ask questions about the vacuum without specifying the scale. 
If the \cc\ arises due to the fluctuations in the energy density of the vacuum, then one needs to understand the structure of the quantum gravitational vacuum at cosmological scales. 
 If the spacetime has a cosmological horizon which blocks information, the natural scale is provided by the size of the horizon,  $L_\Lambda$, and we should use observables defined within the accessible region. 
The operator $H(<L_\Lambda)$, corresponding to the total energy  inside
a region bounded by a cosmological horizon, will exhibit fluctuations  $\Delta E$ since vacuum state is not an eigenstate of 
{\it this} operator. A rigorous calculation (see the first reference in \cite{cc2}) shows that 
  the fluctuations in the energy density of the vacuum in a sphere of radius $L_\Lambda$ 
 is given by 
 \begin{equation}
 \Delta \rho_{\rm vac}  = \frac{\Delta E}{L_\Lambda^3} \propto L_P^{-2}L_\Lambda^{-2} 
 \label{final}
 \end{equation}
 The numerical coefficient will depend on $c_1$ as well as the precise nature of infrared cutoff 
 radius;
 but it is a fact of life that a fluctuation of magnitude $\Delta\rho_{vac}\simeq H_\Lambda^2/G$ will exist in the
energy density inside a sphere of radius $H_\Lambda^{-1}$ if Planck length is the UV cut off. 
On the other hand, since observations suggest that there is a $\rho_{vac}$ of similar magnitude in the universe it seems 
natural to identify the two. Our approach explains why there is a \textit{surviving} cosmological constant which satisfies 
$\rho_{_{\rm DE}}=\sqrt{\rho_{_{\rm IR}}\rho_{_{\rm UV}}}$.
  
Such a  computation of energy fluctuations is completely meaningless in the  models of gravity in which the metric couples to the bulk energy density. Once a UV cut-off at Planck scale is imposed, one will always get a bulk contribution $\rho_{UV}\approx L_P^{-4}$ with  the usual problems. It is only because we have a way of decoupling the bulk term  from contributing to the dynamical equations that, we have a right to look at the subdominant term $L_P^{-4}(L_P/L_\Lambda)^2$. Approaches in which the sub-dominant term is introduced by an ad hoc manner are technically flawed since the bulk term cannot be ignored in these usual approaches to gravity.
Getting the correct value of the cosmological constant from the energy fluctuations is not as difficult as understanding why the bulk value  (which is larger
by $10^{120}$!) can be ignored. Our approach provides a natural backdrop for 
ignoring the bulk term --- and as a bonus --- we get the right value for the cosmological 
constant from the fluctuations. Cosmological constant   is small because it is a  quantum relic.

\section{Conclusions}

 The simplest choice for the negative pressure component in the universe is the cosmological constant; other models based on scalar fields (as well as those based on branes etc. which I  have not discussed) do not alleviate the difficulties faced by \cc\  and --- in fact --- makes them worse. 
 The cosmological constant
  is most likely to be a low energy relic of a quantum gravitational effect or principle and its explanation will require a radical shift in our current paradigm.
  
A new approach to gravity described here could provide  a possible broad paradigm
to understand the cosmological constant. 
 The conceptual
  basis for this approach rests on the following logical ingredients.
 I have shown that it is impossible to solve the \cc\ problem unless the gravitational sector of the theory is invariant under the shift $T_{ab}\to T_{ab}+\lambda_mg_{ab}$. Any approach which does not address this issue cannot provide a comprehensive solution to the \cc\ problem.
 But general covariance requires us to use the measure $\sqrt{-g}d^Dx$ in D-dimensions
in the action. This will couple the metric (through its determinant) to the matter sector. Hence, as long as we insist on metric as the fundamental variable describing gravity, one cannot address
this issue.  So we need to introduce some other degrees of freedom and an effective action which, however, is capable of constraining the background metric.

An action principle, based on the  normalized vector fields in spacetime, satisfies all these criteria mentioned above. The new action does not couple to the bulk energy density and maintains invariance under the shift $T_{ab}\to T_{ab}+\lambda_mg_{ab}$. What is more, the on-shell value of the action is related to the entropy of horizons showing the relevant degrees of freedom scales as the area of the bounding surface.
Since our formalism ensures that the bulk energy density does not contribute to gravity --- and only because of that --- it makes sense to compute the next order correction due to fluctuations in the energy density. This is impossible to do rigorously with the machinery available but a plausible case can be made as how this will lead to the correct, observed, value of the \cc.

An effective theory can capture the relevant physics at 
 the long wavelength limit using  the degrees of freedom contained in the fluctuations of the normalized vectors. 
The  
resulting theory  is  more general than Einstein gravity since the thermodynamic interpretations should transcend classical considerations and incorporate some of the microscopic corrections. 
Einstein's equations provide the lowest
order description of the dynamics and \textit{calculable}, higher order,
corrections arise as we probe smaller scales.    
 The mechanism for ignoring the bulk \cc\ is likely to survive quantum gravitational corrections which are likely to bring in additional, higher derivative, terms to the action.

 \end{document}